\documentclass[reprint,amsmath,amssymb,aps,prl]{revtex4-2}
\usepackage{amsmath}
\usepackage{xcolor}
\usepackage{graphicx}% Include figure files
\usepackage{dcolumn}% Align table columns on decimal point
\usepackage{bm}% bold math
\usepackage{orcidlink}

\begin{document}

\title{Breakdown of hydrodynamics in a one-dimensional cold gas}
\author{{Taras Holovatch}\orcidlink{0009-0005-2953-8730}$^{1,2}$, 
	{Yuri Kozitsky}\orcidlink{0000-0002-4320-8835}$^3$, 
	{Krzysztof Pilorz}\orcidlink{0000-0003-2596-3260}$^3$,
	{Yurij Holovatch}\orcidlink{0000-0002-1125-2532}$^{1,2,4,5}$
} 
\affiliation{
	$^1${Yukhnovskii Institute for Condensed Matter Physics of the National Academy of Sciences of Ukraine}, {Lviv}, {79011}, {Ukraine}\\
	$^2${$\mathbb{L}^4$ Collaboration and Doctoral College for the Statistical Physics of Complex Systems}, {Lviv-Leipzig-Lorraine-Coventry}, {Europe}\\
	$^3${Institute of Informatics and Mathematics, Maria Curie-Sk\l odowska University}, {20-031} {Lublin}, {Poland}\\
	$^4${Centre for Fluid and Complex Systems, Coventry University}, {Coventry} {CV1 5FB}, {UK}\\
	$^5${Complexity Science Hub}, {1030} {Vienna}, {Austria}
}
%\pacs{}
%\date{\today}

\begin{abstract}

The following model is studied analytically and numerically: point particles with masses $m,\mu,m, \dots$ ($m\geq\mu$) are distributed over the positive half-axis. Their dynamics is initiated by giving a positive velocity to the particle located at the origin; in its course the particles undergo elastic collisions. 
We show that, for certain values of $m/\mu$, starting from the initial state where the particles are equidistant the system evolves in a hydrodynamic way:  (i) the rightmost particle (blast front) moves as $t^{\delta}$ with $\delta < 1$;               
(ii) recoiled particles behind the front enter the negative half-axis; (iii)  the splatter -- the particles with locations $x\leq 0$ -- moves in the ballistic way and eventually takes over the whole energy of the system. These results agree with those obtained in S. Chakraborti et al, SciPost Phys. 2022, 13, 074, for $m/\mu=2$ and random initial particle positions.  At the same time, we explicitly found the collection of positive numbers $\{\mathcal{M}_i, i \in \mathbf{N}
\}$ such that, for $m/\mu = \mathcal{M}_i$, $i\leq 700$, the following holds: (a) the splatter is absent; (b) the number of simultaneously moving particles is at most three; (c) the blast front moves in the ballistic way. However, if, similarly as in S. Chakraborti et al, the particle positions are sampled from a uniformly distributed ensemble, for $m/\mu = \mathcal{M}_i$ the system evolves in a hydrodynamic way. 
\end{abstract}

\maketitle 

%\section{Introduction}
{\it I. Introduction. --} An important task of contemporary statistical mechanics is to clarify interconnections between the micro- and macroscopic descriptions of a large interacting particle system, see, e.g., \cite{Spohn}.            
This relates to the transport of energy in a cold gas following explosion -- an 
instant injection of energy at its certain point. 
In \cite{Antal,Chakraborti21,Chakraborti22,kumar}, a cold gas with explosion was modeled as a system of resting hard disks      
undergoing elastic collisions with the excited particles. In the actual formulation, this model and the corresponding terminology were first introduced in \cite{Antal}, where the two-dimensional case was also considered. While two- and three-dimensional versions of the model are more adequate, their rigorous study at the microscopic level is highly challenging. In the one-dimensional models, the collisions are inevitable even if the disks (rods) are shrunk to points. 
However, such one-dimensional models have a degenerate version: if all the particles have equal masses, the dynamics gets trivial (domino-like), as the number of moving particles is conserved  ($=1$) and the blast front evolves as ${\cal R}(t)=v_0 t$. To get rid of this degeneracy, one studies  systems of particles with  masses alternating between different values, 
the number of which is two in the simplest case. Such a model is known as the
alternating hard particle (AHP) gas, see also \cite{Dhar,Garrido,Grass}. In the symmetric version of the AHP gas, the blast fronts move in both directions, and their distance to the injection site evolves as ${\cal R}(t) \sim t^{2/3}$, which was obtained by molecular dynamics (MD) simulations \cite{Chakraborti21} in agreement with the corresponding hydrodynamic result obtained by solving Euler's equation. Another version of the AHP gas was studied in 
 \cite{Chakraborti22}. It is the system of resting point particles distributed over the positive part of the real axis, in which the dynamics is initiated by giving positive velocity to the particle located at the origin. Then the blast front moves in the positive direction whereas recoiled particles behind it can move in the opposite direction and enter the negative half-axis forming a `splatter'. In the sequel, the models of \cite{Chakraborti21} and \cite{Chakraborti22} will be called symmetric and asymmetric AHP gas models, respectively. 	

 Similarly as in \cite{Chakraborti21}, the results of \cite{Chakraborti22} were obtained by making analytic predictions based on solving the corresponding hydrodynamic equation confronted with numerical simulation results of solving Newtonian equations of motion. In the simulations, the initial positions of the particles were 
 sampled from the uniformly distributed ensemble. Among the results of \cite{Chakraborti22} obtained for the system of particles with masses $m,\mu, m, \dots $, $m/\mu=2$, we mention the following: (i) the blast front evolves as ${\cal R}(t)\sim t^{\delta}$ with $\delta \simeq 0.628$; (ii) the splatter moves in the ballistic way; (iii) the energy 
$\mathcal{E}_{x\geq 0}( t )$ of the particles with locations $x\geq 0$ at time $t$  decays to zero as $t^{-\beta}$ with $\beta \simeq 0.1161$; (iv) the number of collisions up to time $t$ increases as $\mathcal{C}(t) \sim t^{2(2-\beta)/3}$ with $\beta$ as just mentioned. It was claimed that essentially the same results were obtained also for $m/\mu=3$ and $m/\mu=10$. 

In this work, we study the model of \cite{Chakraborti22} with various values of
 $m/\mu$. We explicitly found the collection of positive numbers $\{\mathcal{M}_i: i \in \mathbf{N}\}$, see \eqref{M} below, such that, for $m/\mu =\mathcal{M}_i$, $i\leq 700$, the following holds. If the particles are initially located equidistantly, then: (a) the splatter is absent; (b) the number of simultaneously moving particles is at most three; (c) the blast front moves in the ballistic way -- in contrast to
the hydrodynamic picture drawn in \cite{Chakraborti22}. This conclusion is applicable also to the symmetric  AHP gas studied in \cite{Chakraborti21}.   
For $m/\mu$ close to $\mathcal{M}_i$, the front motion remains ballistic-like -- at least in the window the span of which depends on $i$ and $|\mathcal{M}_i-m/\mu|$. At the same time, for other values of $m/\mu$, including $m/\mu=2$, as well as for $m/\mu=\mathcal{M}_i$ and random initial particle positions, our results are essentially the same as those found in \cite{Chakraborti22}. Further discussion of these issues is given below. 

%\section{The Results}
%\subsection{Posing}
{\it II. The Results. Posing. --} If point particles $a$ and $b$,  with masses $m_a$ and $m_b$, undergo an elastic collision, their velocities after, $v'_a$ and $v'_b$, and before, $v_a$ and $v_b$, the collision satisfy
\begin{equation}
 \label{RV}
 v'_a = \frac{m_a - m_b}{m_a+m_b}v_a + \left(1-\frac{m_a - m_b}{m_a+m_b} \right)v_b,
\end{equation}
with $v'_b$ obtained by interchanging $a$ and $b$ in \eqref{RV}.

We consider an infinite system of initially resting point particles numbered $0,1,2, \dots$ with initial positions $x^0_k = k$ and  
masses $m,1, m, 1,\dots$, $m\geq 1$. At time $t=0$, particle 0 (with mass $m$) receives rightward velocity $v_0=1$, and the entire system is gradually set into motion in which the particles undergo elastic collisions and can enter the negative half-axis. 
Their dynamics is observed in the window containing $N$ first particles, i.e., it is terminated 
at $t=t_{\rm{fin}}$ when particle $N-1$ is hit for the first time. 
The following observables are to be considered: $\mathcal{R}(t)$ - the position of the rightmost moving particle at time $t$ (blast front); $\mathcal{C}(t)$ - the number of collisions up to time $t$; 
$\mathcal{E}_{x\geq 0}(t)$ - the energy of particles with coordinates $x\geq 0$ at time $t$.
We will also consider $\mathcal{C}_{\rm fin}= \mathcal{C}(t_{\rm{fin}})$ and $\mathcal{E}_{\rm fin}= \mathcal{E}_{x\geq 0}(t_{\rm{fin}})$
 --
the total observed number of collisions and the energy of all particles with $x\geq 0$  
at $t=t_{\rm{fin}}$. 
The numerical values of the observables in question are obtained on the base of \eqref{RV} by means of the MD simulations performed by filling heap data structure \cite{Williams} with all collisions, potentially possible at the moment of appending, further choosing the most immediate collision still valid after previously processed collisions.  
Typically, the number of interacting particles ranged between $10^4-10^5$ and is indicated below. 
The simulations results are then used to get the time dependence of the observables as powers of time with the help of our method presented in \cite{Holovatch25}. In particular, 
for an observable, $\mathcal{O}$, the $\ln t$-dependence of $\ln \mathcal{O}$ is fitted as a straight 
line with the slope, $\alpha$, in which case  
we write $\mathcal{O}(t) \sim t^\alpha$. For the aforementioned observables, similarly as in \cite{Chakraborti22} we use the exponents
\begin{equation}
 \label{Obs}
 \mathcal{R}(t) \sim t^\delta, \quad \mathcal{E} _{x\geq 0}(t) \sim t^{-\beta}, \quad \mathcal{C}(t) \sim t^\eta.
\end{equation}
In the hydrodynamic regime, they are related to each other by
\begin{equation}
 \label{Rule}
 \eta = 2 \delta , \quad \delta= (2-\beta)/3, 
\end{equation}
with $\beta = 0.11614\dots$, see \cite{Chakraborti22}.
 
\begin{figure}[htb]
	\setlength{\abovecaptionskip}{0pt}
	\setlength{\belowcaptionskip}{0pt}
	\includegraphics[width=1\linewidth]{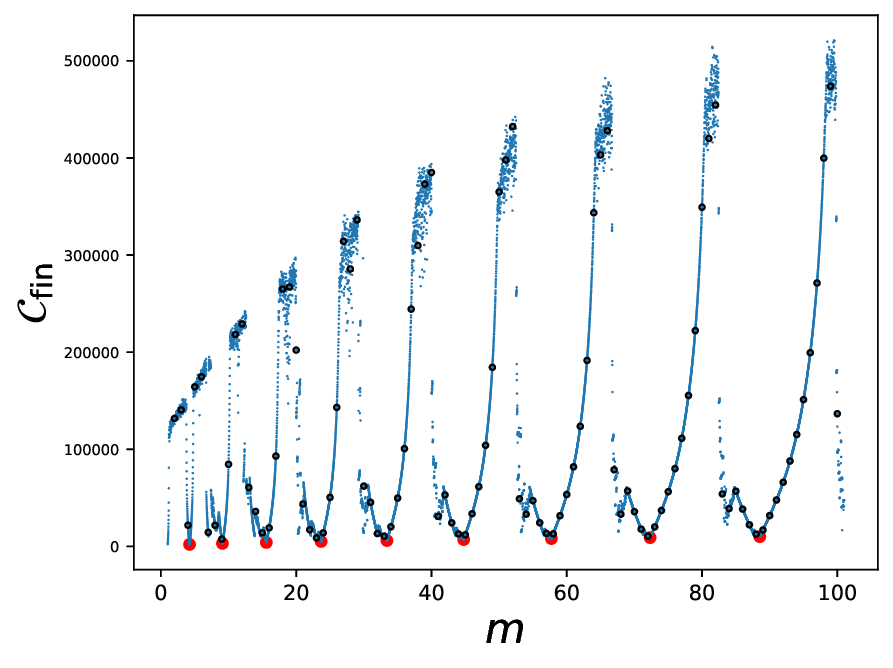}
	\caption{The total number of collisions $\mathcal{C}_{\rm fin}$ as a function of $m$. The scattered blue dots correspond to the resolution $\Delta m=0.01$; black circles correspond to integer values of $m$; red disks correspond to $m=\mathcal{M}_i$. \label{fig1}}
\end{figure}

\begin{figure}[htb]
	\setlength{\abovecaptionskip}{0pt}
	\setlength{\belowcaptionskip}{0pt}
	\includegraphics[width=1\linewidth]{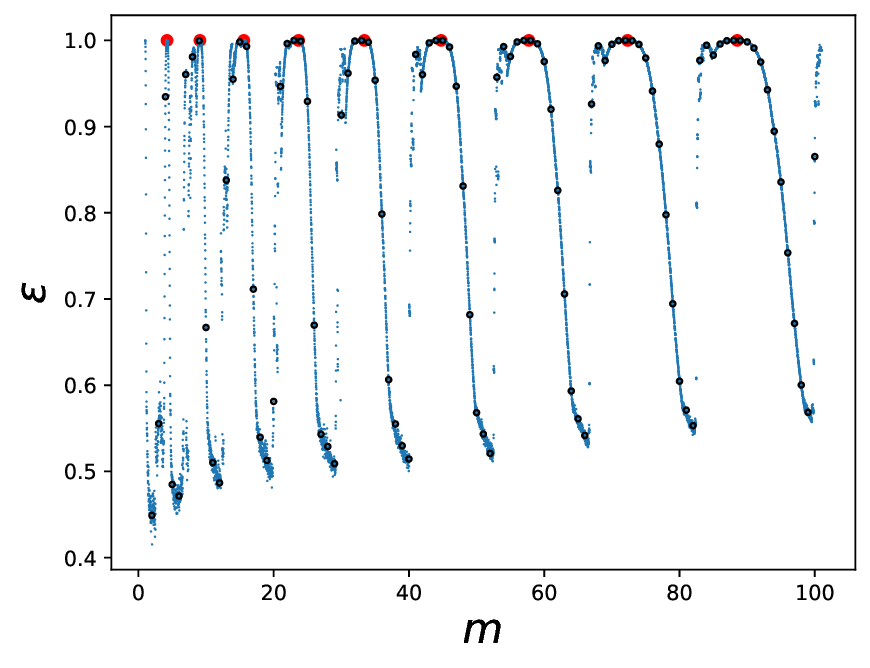}
	\caption{The normalized energy $\varepsilon(m) =2\mathcal{E}_{\rm fin}/m$ as a function of $m$. 
		 Symbols are as in Fig. \ref{fig1}. Red discs correspond to  $\varepsilon(\mathcal{M}_i)=1$.
		\label{fig2a}}
\end{figure}

%\subsection{Observations}
{\it Observations. --}
Our first observation is about the $m$-dependence of the total number of collisions $\mathcal{C}_{\rm fin}(m)$. We present it
in Fig. \ref{fig1} calculated for $N=1,000$ with the step $\Delta m = 0.01$. For particular values of $m$ and $N\geq10,000$ it is also presented in the third column of Table \ref{tab1}. The significant feature of $\mathcal{C}_{\rm fin}(m)$ is that it
periodically attains minima at certain points, $m=\mathcal{M}_i$, with the increasing amplitude of oscillations. 
Additionally, in Fig. \ref{fig2a} we plot  
the $m$-dependence of the normalized energy $\varepsilon (m)= 2\mathcal{E}_{\rm fin}/m$.
Our second observation is  that $\varepsilon(\mathcal{M}_i)=1$, i.e., $\mathcal{E}_{\rm fin}$ is equal to the energy originally injected to the system. Hence, all the particles never appear in the negative half-axis.
Both these observations clearly indicate that, for $m=\mathcal{M}_i$, the system dynamics gets essentially different from that obtained in \cite{Chakraborti22} for $m=2$. Our aim is to study this phenomenon. 

%\subsection{The phenomenon}
{\it The phenomenon. --}  First we consider the dynamics of the triplet  of particles $0,1,2$. 
Let $v^{(1)}_j$, $j=0,1,2$, be their velocities after the first round of mutual collisions, i.e, right after the collisions $0\rightarrow 1$ and $1\rightarrow 2$.  According to \eqref{RV} we then get 
\begin{gather}
 \label{a}
 v_0^{(1)} = \theta, \quad v_1^{(1)} = -\theta (1+\theta), \quad  
  v_2^{(1)} = 1- \theta^2,   
 \end{gather}
where we use
$\theta = (m-1)/(m+1)$, cf. \eqref{RV}.
If $m=\mathcal{M}_1 := 1$, then $\theta = \theta_1:=0$; hence, $v_0^{(1)} = v_1^{(1)} = 0$ and $v_2^{(1)} = 1$, which corresponds to the aforementioned degeneracy. For $\theta >\theta_1$, one gets $v_0^{(1)} >0$, $v_1^{(1)} < 0$, and thus the collision $0\leftrightarrow 1$ between particles 0 and 1 (moving towards each other) occurs at time ${t^{(2)}_{0\leftrightarrow 1}} = 1+ (1+\theta)/\theta(2+\theta)$ and point ${\xi^{(2)}_{0\leftrightarrow 1}} = 2 - 1/(2+\theta)$. Here and below, by $t^{(i)}_{a \leftrightarrow b}$ and $\xi^{(i)}_{a \leftrightarrow b}$ we denote time and position of the collision between particles $a$ and $b$ in round $i$. Thereafter, the velocities change to
$v_0= \theta (\theta^2 + \theta -1)$, $v_1= \theta (1+\theta)^2$, $v_2 = 1-\theta^2$. For $t\geq t^{(2)}_{0\leftrightarrow 1}$, the particle positions are 
\begin{gather}
 \label{d}
 x_0 (t)= 2 - \frac{1}{2+\theta} + \theta (\theta^2 + \theta -1)(t-t^{(2)}_{0\leftrightarrow 1}),\\ \nonumber x_1 (t) = 2 - \frac{1}{2+\theta} + \theta (1+\theta)^2(t-t^{(2)}_{0\leftrightarrow 1}),\\ \nonumber x_2 (t) = 2 + (1-\theta^2) ( t - 1 - 1/(1+\theta)). 
\end{gather}
For small $\theta$, particle $2$ escapes from 1 and hits 3. For $\theta > \sqrt{2}-1$, one has $\theta^2 +2 \theta -1>0$ and then
$v_1 > v_2$. If this holds, the collision $1 \rightarrow 2$ gets possible before $2 \rightarrow 3$ provided it happens at $\xi^{(2)}_{1\leftrightarrow 2} < 3$. The time of $1 \rightarrow 2$ can be found by solving  
$x_1(t) = x_2(t)$, which yields 
\[
 t^{(2)}_{1 \leftrightarrow 2} = 1 + \frac{\theta (2+\theta)}{(1+\theta)(\theta^2 +2 \theta -1)}.
\]
Then 
\begin{equation}
 \label{e}
 \xi^{(2)}_{1\leftrightarrow 2} = x_1 (t^{(2)}_{1 \leftrightarrow 2}) =x_2 (t^{(2)}_{1 \leftrightarrow 2}) = 2 + \frac{1-\theta}{\theta^2 +2 \theta -1}< 3,
\end{equation}
where the latter inequality holds for $\theta > (\sqrt{17} - 3)/2 \simeq 0.562$ which we suppose to hold. 
Thus, the second round of the mutual collisions $0\leftrightarrow 1$, $1\rightarrow  2$ takes place, after which the velocities $v^{(2)}_j$ take the form      
\begin{gather}
 \label{b}
 v^{(2)}_0 = \theta w_2(\theta), \quad v^{(2)}_1 = - (1+\theta)^2 w_2(\theta), \\ \nonumber v^{(2)}_2 = 1- [w_2(\theta)]^2 , \quad {\rm where} \quad w_2(\theta)= \theta^2 + \theta -1. 
\end{gather}
Let $\theta_2 = (\sqrt{5} -1)/2\simeq 0.612$ be the positive root of $w_2$. For $\theta = \theta_2$, one has $m= \mathcal{M}_2 =  \sqrt{5}  + 2$, and also $v_0^{(2)} = v_1^{(2)} = 0$ and $v_2^{(2)} = 1$. Thereafter, particles $2,3, 4$ repeat the dynamics as just described. Then for $m=\mathcal{M}_2$, the number of moving particles does not exceed three and $\mathcal{E}_{x\geq 0} (t) = m/2$ for all $t$. Moreover, the rightmost particle moves with a velocity $v\geq 1 - \theta_2^2$, which then yields the ballistic way of the blast front propagation as $\mathcal{R}(t) \geq  (1 - \theta_2^2)t$. 

For $\theta >\theta_2$, one gets $v_0^{(2)} >0$, $v_1^{(2)} < 0$, and thus particles 0 and 1 will collide once again. Assuming that the next collision will be $1 \rightarrow 2$ not $2\rightarrow 3$ (checked a posteriori), one obtains
\begin{gather}
 \label{c}
 v^{(3)}_0 = w_2(\theta) w_3(\theta), \quad v^{(3)}_1 = - \theta(1+\theta)(2+\theta) w_3(\theta), \\ \nonumber v^{(3)}_2 = 1- w(\theta) w_3(\theta) , \quad w_3(\theta)= \theta^3 +2 \theta^2 - \theta -1, 
\end{gather}
where $w$ is a polynomial. The polynomial $w_3$ has the root $\theta_3 \in( \theta_2, 1)$ such that $w_3(\theta) > 0$ for $\theta > \theta_3$. Then for $m=\mathcal{M}_3 = (1+\theta_3)/ (1-\theta_3) \simeq 9.09783$, one has $v_0^{(3)} = v_1^{(3)} = 0$ and $v_2^{(3)} = 1$. Thus, particles 0,1,2 perform three rounds of mutual collisions as described above after which the first two particles stop. Thereafter, the triplet 2,3,4 repeats the same three rounds, repeated by particles 4,5, 6 afterwards. 

\begin{figure}[!htb]
	\setlength{\abovecaptionskip}{0pt}
	\setlength{\belowcaptionskip}{0pt}
	\includegraphics[width=1\linewidth]{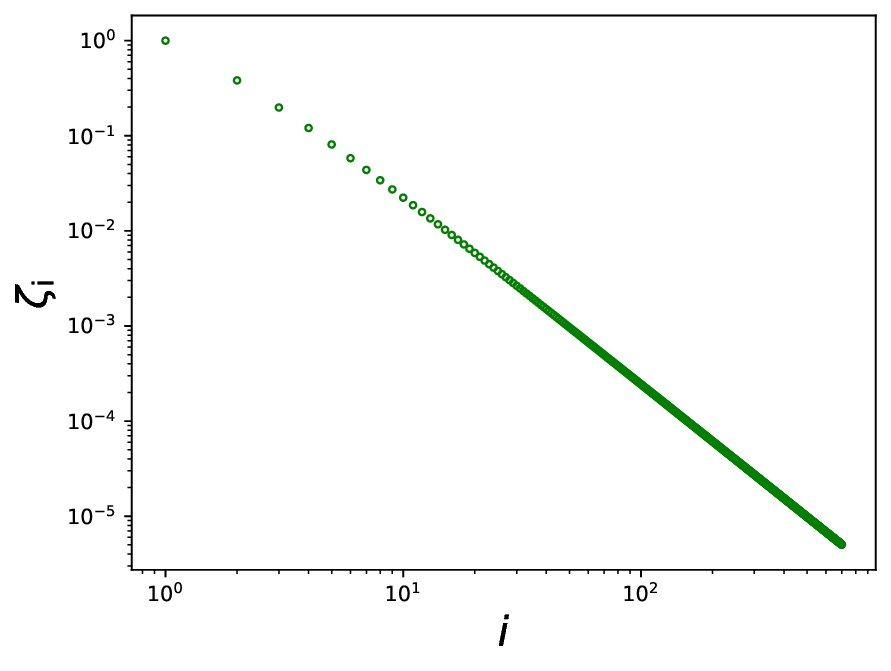}
	\caption{Dependence of $\zeta_i$ on $i\leq 700$, see 
		(\ref{eq_cond}). 
		\label{fig3a}}
\end{figure}

With the help of direct calculations, we found that particles $0,1,2$ perform $i\in \mathbf{N}$ rounds of mutual collisions after which their velocities become $v^{(i)}_0= v^{(i)}_1 =0$ and $v^{(i)}_2=1$ whenever $m=\mathcal{M}_i$, where 
\begin{equation}
 \label{M}
 \mathcal{M}_i = \cot\frac{\pi}{2(2i+1)} \cot\frac{\pi}{2i+1}, 
\end{equation}
which corresponds to $\mathcal{M}_i \sim 8 i^2/\pi^2$. 
By means of symbolic computations we proved that, for $i\leq 700$, the following holds 
\begin{equation}\label{eq_cond}
\zeta_i:=3 - \xi^{(i)}_{1\leftrightarrow 2} > 0 ,
\end{equation}
cf. \eqref{e}, where $\xi^{(i)}_{1\leftrightarrow 2}$ is the point of the last collision $1\leftrightarrow 2$ for $m=\mathcal{M}_i$. It ensures that particles 0,1,2 undergo $i$ rounds of mutual collisions inside the triplet after which $v^{(i)}_0 = v^{(i)}_1 = 0$ and $v^{(i)}_2=1$; thereafter, the motion is transferred to particles 2,3,4, and so on.  The results are presented  in Fig. \ref{fig3a}. 
\begin{table}[htb]
	\setlength{\abovecaptionskip}{0pt}
	\setlength{\belowcaptionskip}{0pt}
	\begin{center}
		\tabcolsep1.2mm
		\begin{tabular}{|c|c|c|c|c|c|}
			\hline &&&&&\\[-0.9em]
			$m$ & $N$ &$\mathcal{C}_{\rm fin}$& $\beta$ & $\delta$ & $\eta$ \\			
			\hline &&&&&\\[-0.9em]
			2 & $10^4$ & 13,369,301 & 0.116 & 0.623 & 1.255 \\
			\hline &&&&&\\[-0.9em]
			7 & $10^4$ &2,217,111& 0.084 & 0.519 & 1.292 \\
			\hline &&&&&\\[-0.9em]
			9 & $10^4$ & 524,070&0.003 & 0.897 & 1.775 \\ 
			\hline &&&&&\\[-0.9em]
			9.2 & $10^4$ &544,283& 0.004 & 0.892 & 1.769 \\ 
			\hline &&&&&\\[-0.9em]
			10 & $10^4$ &21,976,938& 0.115 & 0.621 & 1.254 \\
			\hline &&&&&\\[-0.9em]
			12 & $10^4$ &23,609,099& 0.119 & 0.628 & 1.256 \\
			\hline &&&&&\\[-0.9em]
			13 & $10^4$ &24,063,895& 0.106 & 0.587 & 1.247 \\
			\hline &&&&&\\[-0.9em]
			14 & $10^4$ &25,572,731& 0.109 & 0.601 & 1.206 \\
			\hline &&&&&\\[-0.9em]
			15 & $10^4$ &1,378,587& 0.019 & 0.724 & 1.573 \\
			\hline &&&&&\\[-0.9em]
			16 & $10^4$ &2,930,570& 0.081 & 0.649 & 1.292 \\
			\hline &&&&&\\[-0.9em]
			34 & $10^4$ &1,928,431& 0.019 & 0.741 & 1.584 \\
			\hline &&&&&\\[-0.9em]
			35 & $10^4$ &38,663,827& 0.122  & 0.582 & 1.222 \\
			\hline 
			\hline &&&&&\\[-0.9em]
			9 & $10^5$ &161,666,977& 0.1800 & 0.3021 & 1.0413 \\ 
			\hline &&&&&\\[-0.9em]
			15 & $3\cdot10^4$ &24,068,759& 0.1437 & 0.3769 & 1.1334 \\
			\hline 
		\end{tabular}
	\end{center}
	\caption{\label{tab1}Simulation results for different $N$ and $m$ for the observables as in (\ref{Obs}).
	}
\end{table}

\begin{figure}[htb]
	\setlength{\abovecaptionskip}{0pt}
	\setlength{\belowcaptionskip}{0pt}
	\includegraphics[width=1\linewidth]{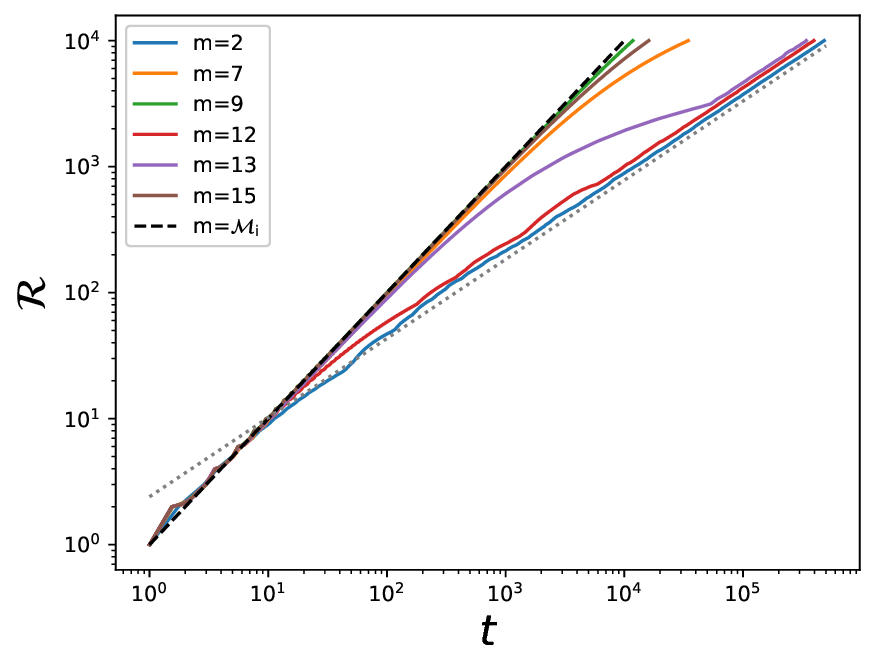}
	\caption{Blast front $\mathcal{R}(t)$ for
		some values of masses $m$, cf \eqref{Obs}. The fitted values of $\delta$  are given in  
		Table \ref{tab1}. Dashed line corresponds to the ballistic evolution with $\delta=1$.
		Dotted line serves as an eye guide to show the hydrodynamics evolution with $\delta \simeq 0.628$. 
		\label{fig5a}}
\end{figure}

In Fig. \ref{fig5a} we show our simulation results for $\mathcal{R}(t)$
at different $m$ and $N=10^4$, where 
the asymptotic ballistic ($\delta=1$) and the hydrodynamic ($\delta\simeq 0.628$) regimes, as well as  
intermediate ones, are presented. 
Clearly, the power-law fitting of the data leads to the size 
dependent values of the (effective) exponents, see \cite{Holovatch25} for more details. Our numerical results are summarized in Table \ref{tab1}. The change in the effective exponents with the increase
of $N$ is especially pronouncing for $m$ close to $\mathcal{M}_i$. For $m=9$, which is close to $\mathcal{M}_3 \simeq 9.09783$, by comparing the  
values of the exponents for $N=10^4$ and $N=10^5$, one can expect that the initially 
ballistic-like dynamics crosses over to a hydrodynamic regime through intermediate 
phases characterized by much slower propagation of the blast front. 
The time of reaching the asymptotic hydrodynamic regime would increase as $m$ approaches 
$\mathcal{M}_i$. At $m=\mathcal{M}_i$, the hydrodynamic regime is never reached. Red discs in Figs. \ref{fig1} and \ref{fig2a} correspond to several first points  $\mathcal{M}_i, i=1...9$, Eq. \eqref{M}.

%\section{Discussion}
{\it III. Conclusions and Discussion. --}
By employing the elastic collision rules (\ref{RV}) we derived the set of positive numbers $\{\mathcal{M}_i: i\in \mathbf{N}\}$, see \eqref{M}, such that for the asymmetric AHP gas with particles initially located at $x^0_k = k=0,1,2, \dots$ and masses satisfying $m/\mu = \mathcal{M}_i$, $i\leq 700$, the blast front dynamics is of the ballistic type, that deviates from the 
hydrodynamic scenario of \cite{Chakraborti21,Chakraborti22}. This conclusion 
applies also to the symmetric AHP gas studied in \cite{Chakraborti21}. In the latter case, the blast front moves rightward as if the left part of the system were absent. At the same time, for the asymmetric AHP gas with $m/\mu = \mathcal{M}_i$ and the initial particle positions sampled from a uniformly distributed ensemble, the blast front dynamics is essentially the same as that of \cite{Chakraborti22}. This observation points to the nonergodicity of the asymmetric AHP gas model with $m/\mu = \mathcal{M}_i$, $i\leq 700$, understood as the dependence on the initial state preserved at all moments of time. The derivation of the formula in \eqref{M}, further details on the MD simulations, as well as an extended discussion of the results presented here will be done in our forthcoming paper.

{\it Data availability statement. --} Data used to produce Figs. \ref{fig1}, \ref{fig2a} and \ref{fig3a} are available at \href{https://github.com/tholovatch/Papers-data}{https://github.com/tholovatch/Papers-data.} Data used to produce Fig. \ref{fig5a} and Table \ref{tab1} are available from the corresponding author, [T. H.], upon reasonable request.

\end{document}